\documentclass[useAMS,usenatbib]{mn2e}
\usepackage{amsmath}
\usepackage{graphicx}
\usepackage{color}
\usepackage{natbib}
\usepackage{epstopdf}

\newcommand{\be}{\begin{equation}}
\newcommand{\ee}{\end{equation}}
\newcommand{\bea}{\begin{eqnarray}}
\newcommand{\eea}{\end{eqnarray}}
\newcommand{\beas}{\begin{eqnarray*}}
\newcommand{\eeas}{\end{eqnarray*}}
\newcommand{\bfig}{\begin{figure}}
\newcommand{\efig}{\end{figure}}
\newcommand{\bfigs}{\begin{figure*}}
\newcommand{\efigs}{\end{figure*}}
\newcommand{\bt}{\begin{table}}
\newcommand{\et}{\end{table}}
\newcommand{\drm}{\mathrm{d}}

\def\simlt{\lower.5ex\hbox{$\; \buildrel < \over \sim \;$}}
\def\simgt{\lower.5ex\hbox{$\; \buildrel > \over \sim \;$}}
\def\simpt{\lower.5ex\hbox{$\; \buildrel \propto \over \sim \;$}}

\def\Mpc{\mbox{Mpc}}
\def\cMpc{\mbox{comoving Mpc}}

\def\trace{\mbox{tr}}
\def\Da{\mathcal{D}}
\def\Dc{{D}}

\def\arcsect{\mbox{arcsec}}

\def\msun{\mbox{M}_\odot}

\def\mnras{MNRAS}
\def\apj{ApJ}
\def\apjl{ApJL}
\def\apjs{ApJ Sup.}

\def\aap{AAP}

\def\nat{Nature}

\definecolor{mylabelcolor}{rgb}{0.5,1,1}

\title{GLAMER Part II: Multiple Plane Gravitational Lensing}

\date{\today}

 \author[Petkova et al. 2013] {\parbox{\textwidth}{Margarita
    Petkova$^{1,2,3}$\thanks{E-mail:
       margarita.petkova@lmu.de},R. Benton Metcalf$^1$,       
    Carlo Giocoli$^{1,4,5}$}
   \\ \\ 
   $^{1}$ Dipartimento di Fisica e
   Astronomia, Universit\`a di Bologna, viale Berti Pichat 6/2,
  40127, Bologna, Italy \\
  $^{2}$  Department of Physics, Ludwig-Maximilians-Universit\"at, Scheinerstr. 1, D-81679 M\"unchen, Germany\\
  $^{3}$  Excellence Cluster Universe, Boltzmannstr. 2, D-85748 Garching, Germany\\
  $^{4}$ INAF - Osservatorio Astronomico di
   Bologna, via Ranzani 1, 40127, Bologna, Italy \\ 
  $^{5}$ INFN - Sezione di Bologna, viale Berti Pichat 6/2, 
   40127, Bologna, Italy 
 }

\begin{document}

\maketitle

\begin{abstract}

  We  present an  extension to  multiple planes of the  gravitational
  lensing  code {\small  GLAMER}.  The  method entails  projecting the
  mass  in the  observed light-cone  onto  a discrete  number of  lens
  planes and inverse ray-shooting from  the image to the source plane.
  The mass on each plane can be represented as halos,
  simulation particles, a projected mass map extracted form a
  numerical simulation or any combination of these.  The image
  finding is done in a source  oriented fashion, where only regions of
  interest are iteratively refined on  an initially coarse image plane
  grid.  The calculations are performed in  parallel on shared  memory
  machines.  The code is able to handle different types of analytic halos
  (NFW, NSIE, power-law, etc.),  haloes extracted from  numerical simulations
  and clusters constructed from semi-analytic models ({\small MOKA}).  
  Likewise, there are several different options for modeling the
  source(s) which can be distributed throughout the light-cone.   The
  distribution of matter in the light-cone can be either taken from a pre-existing N-body
  numerical simulations, from halo catalogs,  or are generated from an
  analytic mass  function.  We present  several tests of the  code and
  demonstrate some of its applications such as generating mock 
  images of galaxy and galaxy cluster lenses.

\end{abstract}


\section{Introduction}\label{sec:intro}

Gravitational  lensing  has  become  a tool  of  greater  and  greater
importance  to cosmology  and the  study of  galaxy structure.   Large
scale  weak lensing  surveys are  being used  to measure  cosmological
parameters     and     study     dark     energy\footnote{see     DES:
  http://www.darkenergysurvey.org,                          Pan-Starrs
  :http://pan-starrs.ifa.hawaii.edu}  and larger  ones are  planned in
space       and       with        purpose       built       telescopes
\citep{euclidredbook,anderson01}.  At the same  time strong lensing is
being           used            to           study           cosmology
\citep{2012arXiv1208.6010S,2013arXiv1306.1272T}    and     the    mass
distribution  within galaxies  down  to  unprecedentedly small  scales
\citep{metcalf12,2013arXiv1307.4220X,2012Natur.481..341V}.   Long term
monitoring  of strong  lenses is  being carried  out with  networks of
telescopes    spanning   the    globe    to    obtain   time    delays
\citep{2012A&A...538A..99S,treu2013}.   The   Hubble  Space  Telescope
(HST)     has      spent     many     orbits      observing     lenses
\citep{postman12,coe12,coe13,zitrin13}     and    many     more    are
scheduled.\footnote{HST                Frontier                Fields:
  http://www.stsci.edu/hst/campaigns/frontier-fields/}   Getting   the
most out  of these huge  observational efforts will require  very good
simulations of  the lensing phenomena  for interpretation of  the data
and to control  systematic effects.  These simulations should
include all  the physical and  instrumental effects that we  know will
affect the data.  Currently, weak lensing simulations have limitations
in  resolution and  the contribution  of baryonic  physics is  usually
neglected.  In addition, they often use unperturbed light rays instead
of  fully  tracing their  paths  through  the simulation.   Individual
galaxies and the distortions they experience are not resolved in these
simulations.  Strong  lensing and  cluster lens simulations  often use
highly idealized  lens models and  usually neglect the  environment of
the lens and contributions to the lens that lie at different redshifts
along the  line of sight.  The  {\small GLAMER} code is  an attempt to
improve this situation in a way  that can be applied to many different
lensing situations on different scales.

Typically, simulations of strong  gravitational lensing treat the lens
as  a  single  plane  without  foreground  and  background  structure.
Systematic  studies,   on  clusters  extracted  from   a  cosmological
numerical      simulation,      have       been      conducted      by
\citet{meneghetti10a,meneghetti11}    With   the    aim   of    better
understanding  of   the  properties   of  galaxy  clusters   that  can
potentially act as strong lenses. In these works however the simulated
clusters are analyzed  using the single lens  approximation.  The same
method has  also been  used by \citet{giocoli12,boldrin13},  where the
individual cluster components are analytically modeled. However, it is
probable that  structures along the  line of sight have  a significant
effect on the observed properties of the lens: both in the weak and in
the strong  lensing regime.  For  example, the inconsistencies  in the
flux         ratios         between        four-images         quasars
\citep{metcalf01,dalal02,2002ApJ...567L...5M,2006MNRAS.367.1367A,maccio06b,2009A&A...498...49C,2009MNRAS.398.1235X,2011ApJ...741..117C}
can  be   at  least  partially  explained   by  line-of-sight  objects
\citep{M04b,2009MNRAS.398.1298P,2005ApJ...635L...1W,2010MNRAS.408.1721X}.
Furthermore, in cluster  lensing, the mass and  concentration of halos
derived from lensing  properties can be affected  by the line-of-sight
contribution.  \citet{2009MNRAS.398.1298P} showed that the presence of
uncorrelated large scale  structures can boost the  Einstein radius by
as much as 50\%.

Weak lensing simulations are usually done by creating a shear map on a
fixed resolution  grid via FFT  of the  potential on one  uniform grid
\citep{2000ApJ...530..547J,  2003ApJ...592..699V, 2007A&A...471..731P,
  2013arXiv1309.1161A,        2001MNRAS.327..169H,2009ApJ...701..945S,
  2011ApJ...742...15T} or on  separate grids for long  and short range
\citep{2007MNRAS.382..121H,2009A&A...499...31H}.   The  resolution  of
the grid is much larger than the  image of a typical galaxy.  The shear at the
positions of  individual galaxies is  found by interpolating  off this
grid and is assumed  to be uniform over the image  of the source.  The
lensing  mass in  the light -cone is  projected onto  multiple lensing
planes and rays are traced from  the observer to the source.  Also the
simulation used to represent the  mass distribution typically does not
have high enough resolution to represent the inner regions of halos or
any  of  the  baryonic  physics happening  there.   Because  of  these
limitations no  strong or quasi-strong lensing  of individual galaxies
will  occur  in these  simulations.   Although  the contribution  from
smaller scales  to the weak lensing  might be small it  is a potential
source of  systematic error and it is  important to characterize it  in the
future high precision weak lensing surveys.

To ensure better resolution and less computational demand, other codes
have  used the  tree  force  algorithm \citep{1986Natur.324..446B}  to
calculate     the    deflection     and     distortion    of     rays.
\cite{1998ApJ...494...29W}   used   this   method   particularly   for
simulations of  quasar microlensing.  And  \cite{ 2007MNRAS.376..113A}
use it for lensing by N-body  particles.  Both of these use the single
plane approximation.

Another  set of  codes \citep{1999MNRAS.306..567F,2012MNRAS.420..155K}
use the ray  bundle method, where a collection of  rays is traced from
the image to the observer, integrating the and computing the shear and
convergence  of the  bundle along  the way.    A third  class of  codes
\citep{1999MNRAS.308..180C,2008MNRAS.388.1618C}      calculate     the
distortion to ray-bundles along unperturbed  paths that pass through a
three dimensional  simulated potential  of a  cosmological simulation.
Still others  use  the  unperturbed  paths,   also  know  as  the  the  Born
approximation,  but put  the  mass from  the  simulations on  multiple
planes \citep{2004APh....22...19W,2007MNRAS.382.1494H}     The     Born
approximation  can   cause  systematic  errors  and   is  clearly  not
sufficient in the case of strong lensing.
\begin{figure*}
\centering
\includegraphics[width=0.95\textwidth]{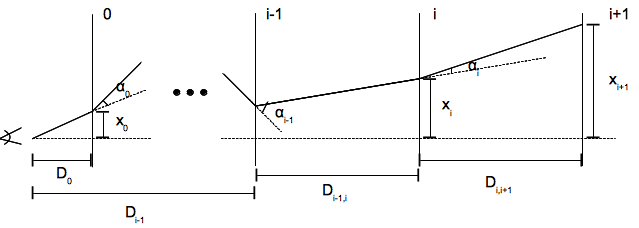}
\caption[]{A schematic view of the multi-plane formalism, as described
  in eqn.~\ref{eq:recursion_x}. \label{fig:scheme}} \end{figure*}

In  \cite{M&P2013} (here  after MP)  we describe  the {\small  GLAMER}
code's structure, its adaptive mesh refinement capabilities and how it
calculates the deflection of light rays by a single lens plane.  These
capabilities  alleviate  many  of  the  limitations  codes  have  with
resolution and make it possible  to construct fully resolved images of
millions  of galaxies  in  a  single simulation.   In  this paper,  we
describe the code's  ability to simulate lensing  in three dimensions;
that is  with the  lensing mass and  also the  sources distributed
throughout the light-cone in angle and redshift.  We also discuss some
of the ways a light-cone can be generated and input into the code.

There  are many  interesting questions  that can  be addressed  with a
lensing code  like the  one we  have developed.   Among them  are: the
importance of line-of-sight objects to strong lenses, the importance of
strongly  lensed  individual galaxies  to  weak  lensing surveys,  the
statistical properties of strong lenses, the importance of multi-plane
effects on precision weak lensing, and many more.

We  start  this   paper  by  presenting  the  basics   of  lensing  in
Sec.~\ref{sec:basics}.   We  elaborate   on  the  multiple  plane
formalism   in   Sec.~\ref{sec:multip_formalism}.   We   continue   in
section.~\ref{sec:implementation} by describing  the implementation of
this    formalism     in    the    {\small    GLAMER}     code.     In
Sec.~\ref{sec:light_cones} we  describe the  different types  of light-cones implemented  in the code.  In  Sec.~\ref{sec:mill_lc} we discuss
the   use a  simulated halo catalog to populate the  light-cone  and   in
Sec.~\ref{sec:massfunc_lc}, the generation of a light-cone from a mass
function.  Then  in Sec.~\ref{sec:tests}  we perform several  tests of
the  code, determining  the optimal  number  of lensing  planes to  be
used.  We also  show some  future applications  of the  code, such  as
Einstein radius mapping of galaxy clusters and producing galaxy-galaxy
strong   lens    catalogs.    We    conclude   our    discussions   in
Sec.~\ref{sec:conclusions}.   For more  extended  information we  have
also included several appendices.


\section{lensing basics}\label{sec:basics}
We will assume  a weak field approximation to  General Relativity.  In
this case, light  propagates through the universe on  the geodesics of
the  perturbed  Robertson-Walker metric.   The  line  element of  this
metric in the Newtonian gauge is \be \mathrm{d}s^2 = a^2\left[ -\left(
  1+\frac{2\Phi}{c^2}\right)\mathrm{d}\tau      ^2       +      \left(
  1-\frac{2\Phi}{c^2}\right)\mathrm{d}l^2 \right],
\label{eqn:metric}
\ee  where  $a$  is  the  cosmological scale  factor,  $\Phi$  is  the
Newtonian  potential,  $c$  is  the  speed of  light,  $\tau$  is  the
conformal time  time, and \bea  \mathrm{d}l^2 &=& \mathrm{d}\chi  ^2 +
\Da^2(\chi)\mathrm{d}\omega    ^2   ,   \\    \mathrm{d}\omega^2   &=&
\mathrm{d}\theta^2+\sin^2 \theta \mathrm{d}\phi^2  .  \eea Here $\chi$
is  the  comoving  distance  and  $\Da(\chi)$,  the  comoving  angular
diameter distance, is defined for different curvatures $K$ as follows:
\bea \Da(\chi) =
\begin{cases}
R \sin( \chi/R) & K=+1, \\
\chi & K=0, \\
R \sinh( \chi/R) & K=-1.
\end{cases}
\eea    where    $R$   is    the    curvature    scale   ($R^{-1}    =
H_o\sqrt{1-\Omega_m-\Omega_\Lambda}/c$  in  the standard  cosmological
model before radiation domination).

The geodesic  equation for a  null geodesic  on small enough  scale to
ignore the curvature $R$ is
\begin{equation}
\frac{\drm^2 {\bf x}}{\drm\lambda^2} = - \frac{2}{c^2} \left[ \nabla \Phi - \left( \frac{\partial {\bf x}}{\partial \lambda } \cdot \nabla \Phi \right) \frac{\partial {\bf x}}{\partial \lambda } \right]
\end{equation}
neglecting higher order terms in  $\Phi$ and $\lambda$ being an affine
distance along the  path.  From the prospective of  the light ray this
can be written more simple as
\begin{equation} \label{eq:geodesic}
\frac{\drm^2 {\bf x}_\perp}{\drm\lambda^2} = - \frac{2}{c^2} \nabla _\perp \Phi\left( {\bf x} \right)  ~~~,~~~ \frac{\drm^2 {\bf x}_\parallel}{\drm\lambda^2} = 0
\end{equation}
where  ${\bf x}_\perp$  and  ${\bf x}_\parallel$  are the  coordinates
perpendicular   and  parallel   to   the   direction  of   propagation
respectively.

If we imagine  a thin slab that the ray  enters perpendicularly we can
integrate equation~(\ref{eq:geodesic}) with  a unperturbed path within
the slab.   We can then calculate  the deflection angle of  the ray on
leaving the slab,
\begin{equation}
{\pmb \alpha}\left({\bf x}_\perp\right) = - \frac{2}{c^2} \nabla _\perp \phi({\bf x}_\perp)  ,
\end{equation}
where $\phi({\bf x}_\perp)$ is now the surface potential.

The surface potential is related to the surface density by 
\begin{equation}
\nabla _\perp^2 \phi({\bf x}_\perp) = 4 \pi G \,\Sigma\left({\bf x}_\perp\right)
\end{equation}
where $\Sigma({\bf x}_\perp)$ is the surface density within the slab
which can be solved to get
\begin{equation}\label{eq:poissons}
\phi({\bf x}) = \frac{4G}{c^2} \int \drm^2x'~ \Sigma({\bf x}') \ln|{\bf x} - {\bf x}'|.
\end{equation}


The position on the sky will be ${\pmb \theta}$.  The position of this
point where  there no  lensing will be  ${\pmb \theta}^S$,  the source
position.  The magnification matrix, or the Jacobian matrix of the map
between these coordinates, is
\begin{equation}\label{eq:A}
{\bf A} \equiv \left[ \frac{\partial \theta^S_i}{\partial \theta_j}\right].
\end{equation}
This matrix  is often  decomposed into the  convergence, $\kappa({\pmb
  \theta})$ and  the components of shear  $\gamma({\pmb \theta})$.  In
the    notation     of    appendix~\ref{sec:ellipt-matr-repr},    this
decomposition is
\begin{equation}\label{eq:Adecomposition}
{\bf A} = (1-\kappa) {\bf I} - \gamma_1{\pmb \sigma}_1 - \gamma_2{\pmb \sigma}_2 - \gamma_3{\pmb \sigma}_3
\end{equation}
We have included a third  component of ``shear'', $\gamma_3$, which is
not usually included, but, as we  will see, is required when there are
more than one lens plane.   The magnification of an infinitesimal point
on the sky is then \be \mu = \frac{1}{{\rm det} {\bf A}}.  \ee

\subsection{single lensing plane}
It is instructive to consider the case of a single lens plane where it
is assumed that the light rays follow unperturbed geodesics outside of
the lens plane.  This is the case usually considered in the literature
so  expressing it  here  in our  notation will  serve  to clarify  our
approach and what new phenomena arise from multi-plane lensing.

In this case, the source and image positions are related by
\begin{equation}
{\pmb \theta}^S = {\pmb \theta} - \frac{\Da_{ls}}{\Da_s}{\pmb \alpha}\left( {\pmb \theta} \right) 
\end{equation}
where  $\Da_s$ is  the angular  diameter  distance to  the source  and
$\Da_{ls}$, between lens plane and  the source (${\bf x}_\perp = {\pmb
  \theta} \Da_l$).   In this case,  the convergence can be  written in
terms of the surface density on the plane
\begin{equation}
\kappa({\pmb \theta}) = \Sigma({\pmb \theta})/\Sigma_{\rm crit} ~~~,~~~~
\Sigma_{\rm crit} \equiv \frac{c^2}{4\pi G} \frac{\Da_s}{\Da_l\Da_{ls}}
\end{equation}
so  in   this  case  $\kappa({\pmb  \theta})$  can   be  considered  a
dimensionless surface density.  This is not the case for multiple lens
planes.  The  deflection field  is a potential  field which  mean that
$\gamma^3 =0$, the field ${\pmb \theta}^S - {\pmb \theta}$ has no curl
and the  shear field  has no  B-modes.  Also every  point on  the lens
plane has a  one-to-one correspondence to a point  on the image plane.
As  we  shall see  these  things are  not  guaranteed  when there  are
multiple lens planes.

\begin{figure*}
\centering
\includegraphics[width=0.33\textwidth]{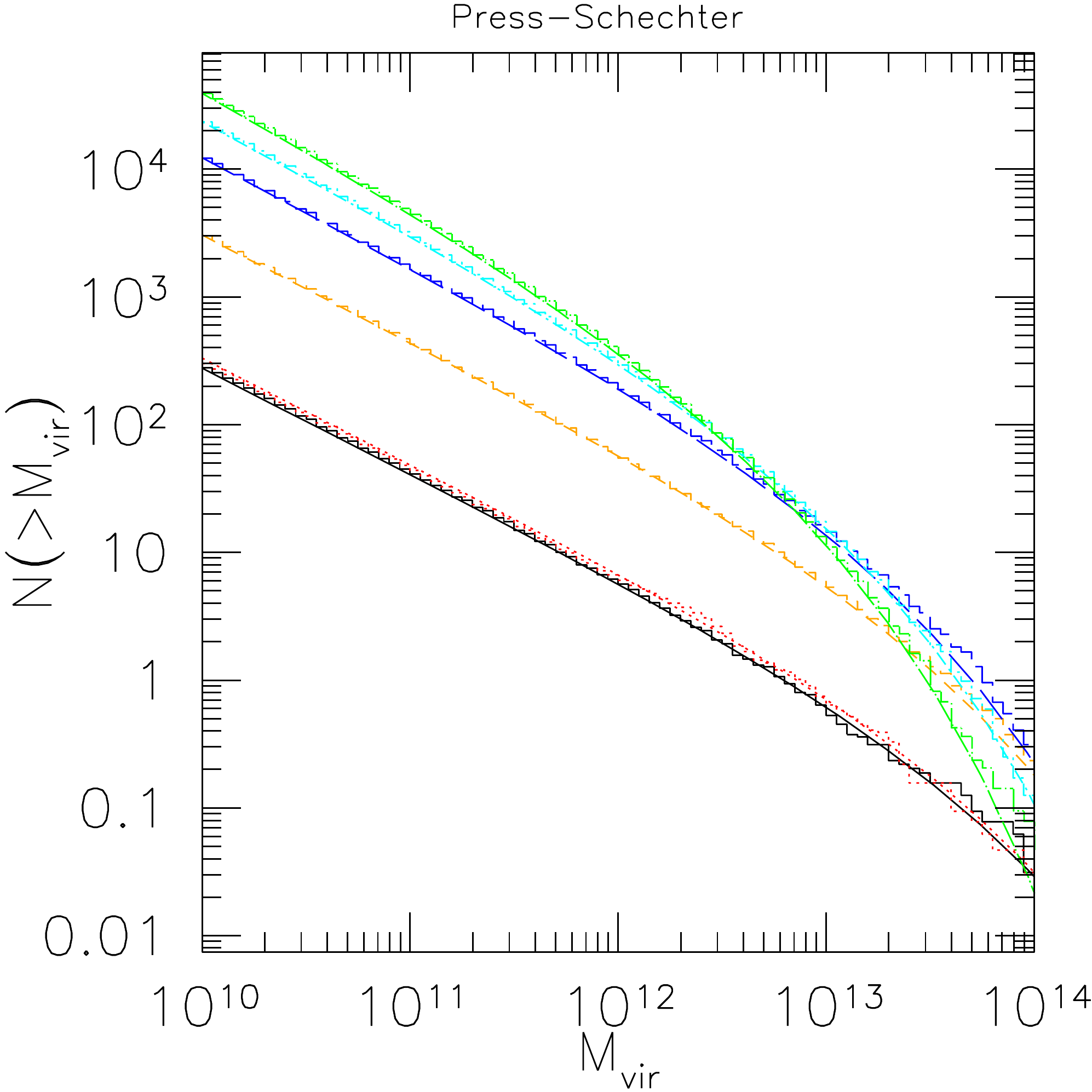}
\includegraphics[width=0.33\textwidth]{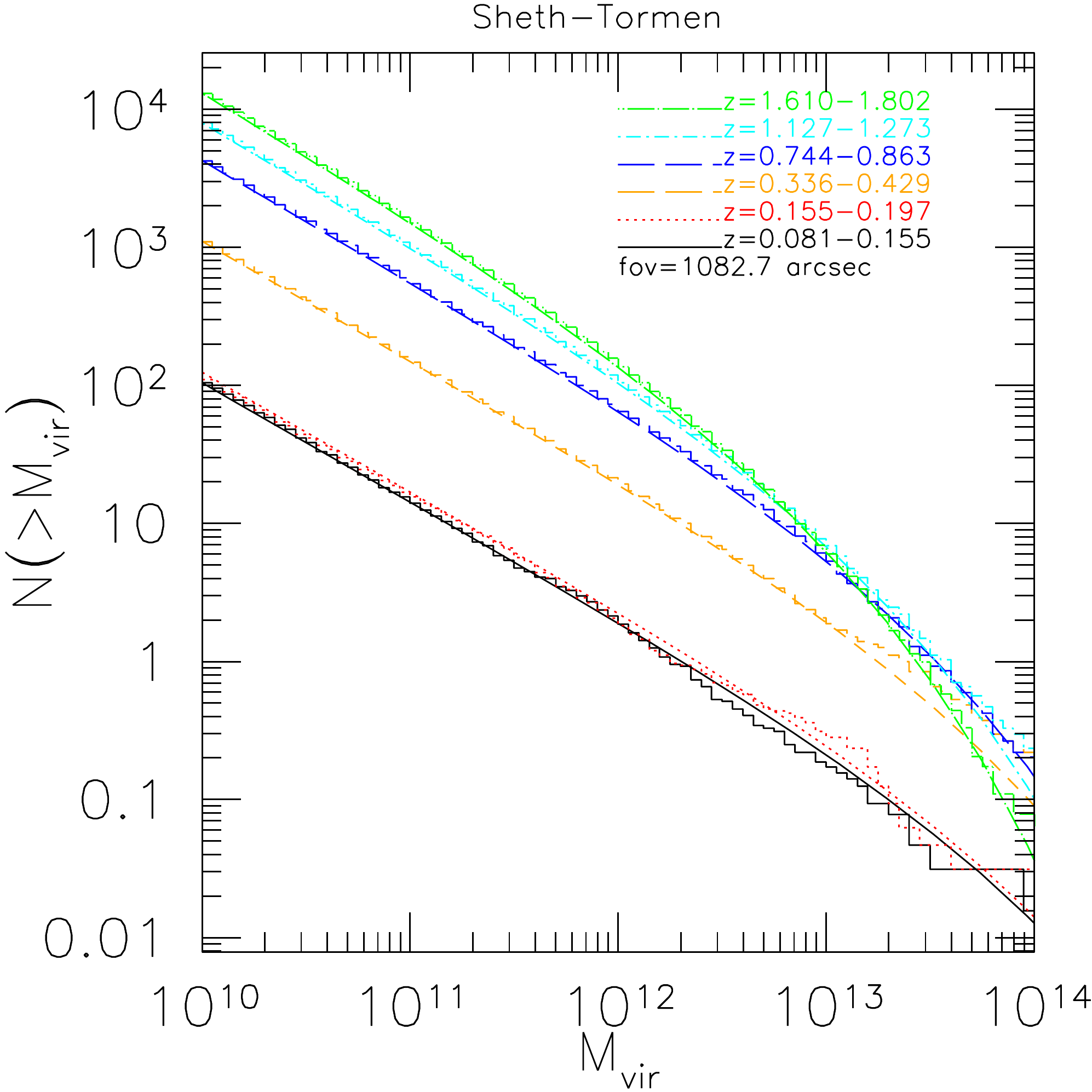}
\includegraphics[width=0.33\textwidth]{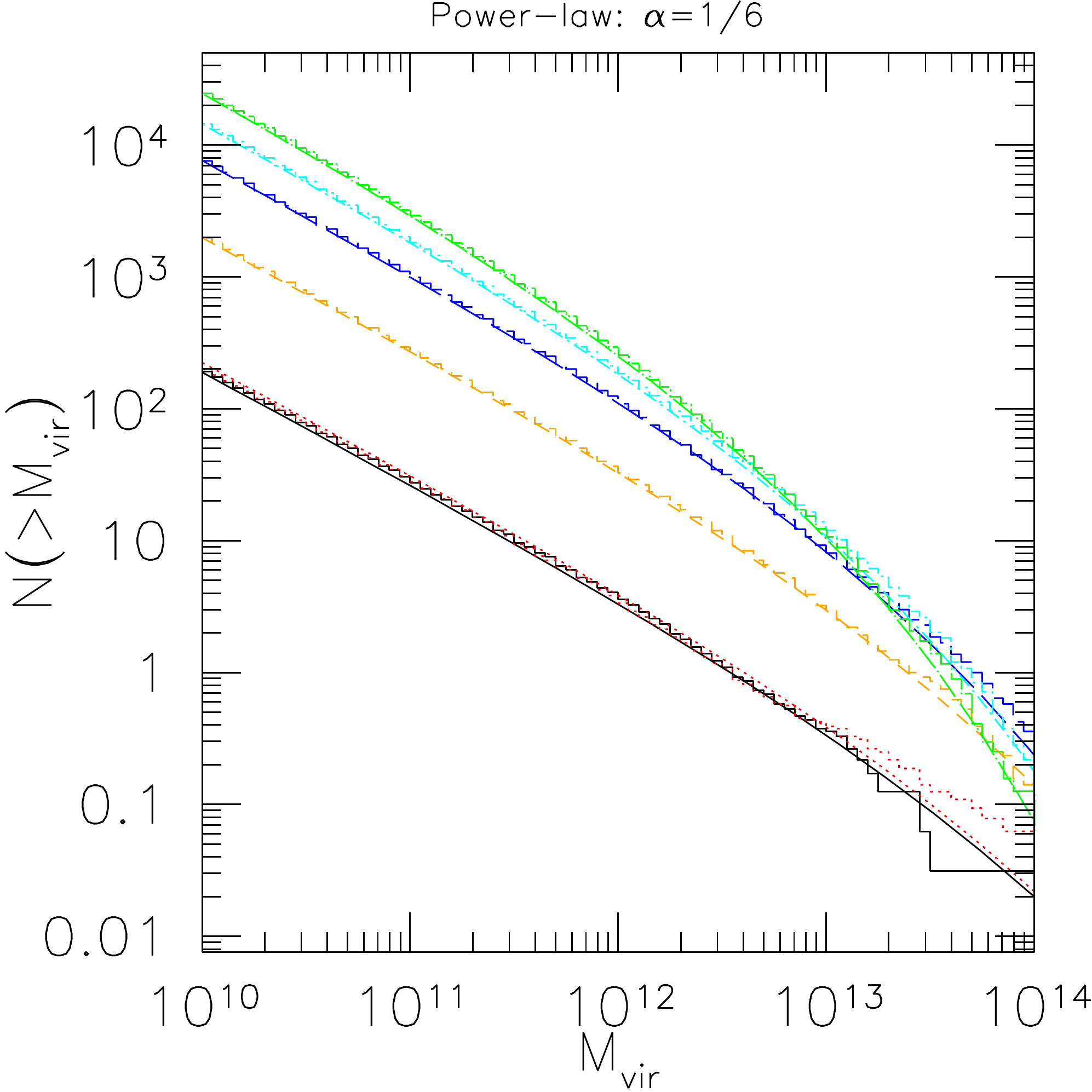}
\caption[]{Halo  mass function,  recreated in  a light -cone of  $10^6
  \arcsect^2$, for different redshift bins.   The left panel shows the
  Press-Schechter mass  function, the  middle one --  the Sheth-Tormen
  one, and  the right one --  a power law  with a slope of  $1/6$. The
  distribution of  masses (histograms), for  which we put  together 64
  different light-cone  realizations, is  in very good  agreement with
  the     theoretical    mass     functions     (lines)    for     all
  redshifts.\label{fig:massF}}
\end{figure*}

\section{multi-plane formalism}\label{sec:multip_formalism}
In the  multi-plane formalism, the  mass in the universe  is projected
onto multiple lensing  planes, perpendicular to the line  of sight. As
light travels from  the lensed object to the  observer, it experiences
deflection from  each plane and  passes unperturbed through  the space
between  the  planes.   The  multiple  plane  lens  equation  has  the
following  form \be  {\pmb \theta}^S  ={\pmb \theta}  - \sum_{i=0}^{N}
\frac{\Da_{is}}{\Da_s} {\pmb \alpha}^i({\pmb \theta}^i) \, ,
\label{eqn:basicM}
\ee where  $N$ is  the number  of planes, $\rm  \Da_s$ is  the angular
diameter distance  to the source  and $\Da_{is}$ -- between  plane $i$
and the source. See figure~\ref{fig:scheme}.

For the  purpose of our  work, from now on  we will use  only comoving
positions  and angular  distances: \bea  {\bf x}  = (1+z)x  ,\\ \Dc  =
(1+z)\Da  ,  \eea   where  $z$  is  the  redshift  of   the  plane  in
consideration.

\begin{figure}
\centering
\includegraphics[width=0.5\textwidth]{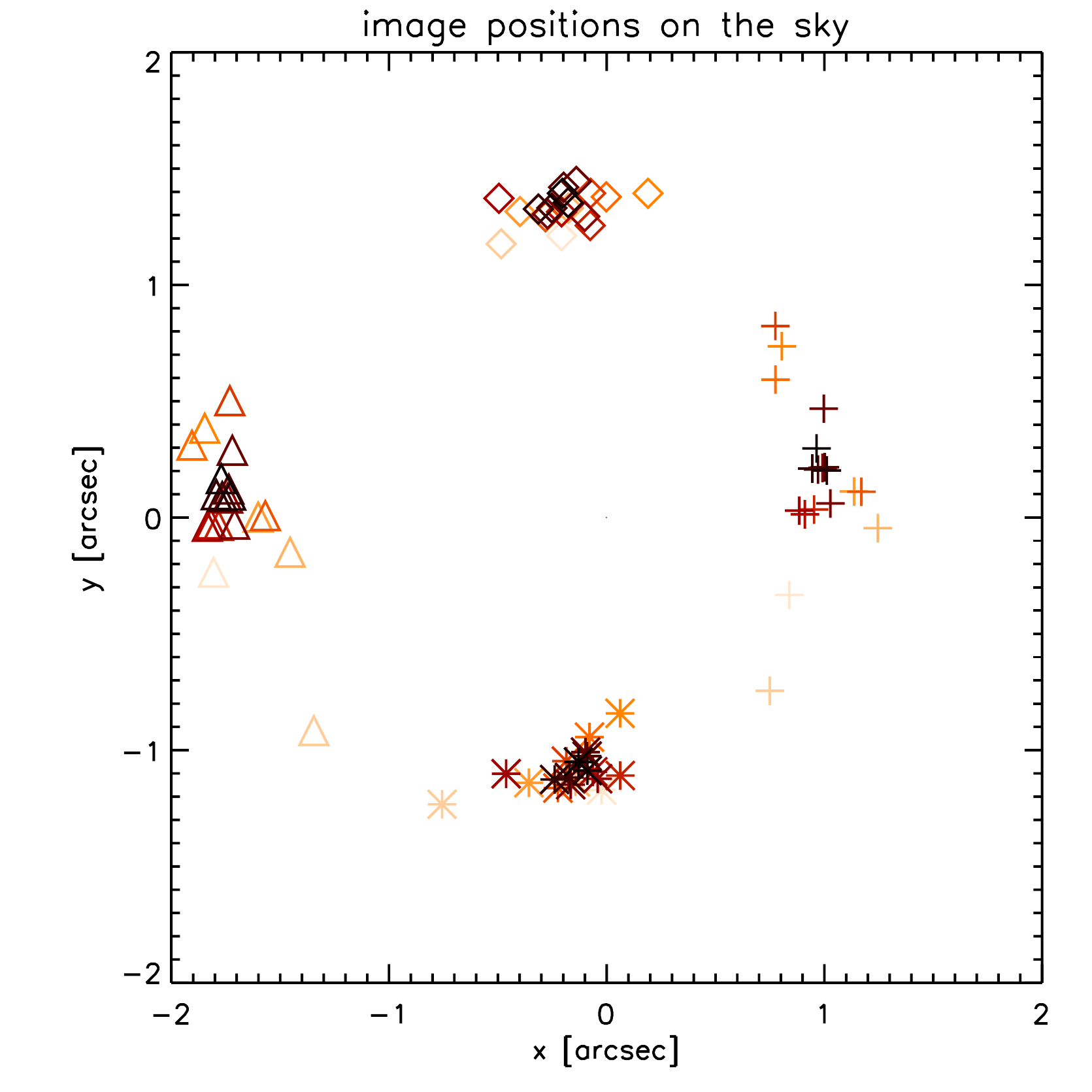}
\caption[]{Position of the four images (different symbols) of a lensed
  quasar  at redshift  $z=3.62$ by  a galaxy  at redshift  $z=0.42$ in
  simulations with different numbers of  lensing planes (6, 8, 12, 14,
  16, 18, 20, 22,  24, 26, 28, 30, 32, 44, 56, 68,  80, 92, 104).  The
  color of the symbols darkens  with increasing number of planes.  The
  positions clearly converge as the number of planes increases.
\label{fig:pos}} 
\end{figure}

\begin{figure}
\centering
\includegraphics[width=0.53\textwidth]{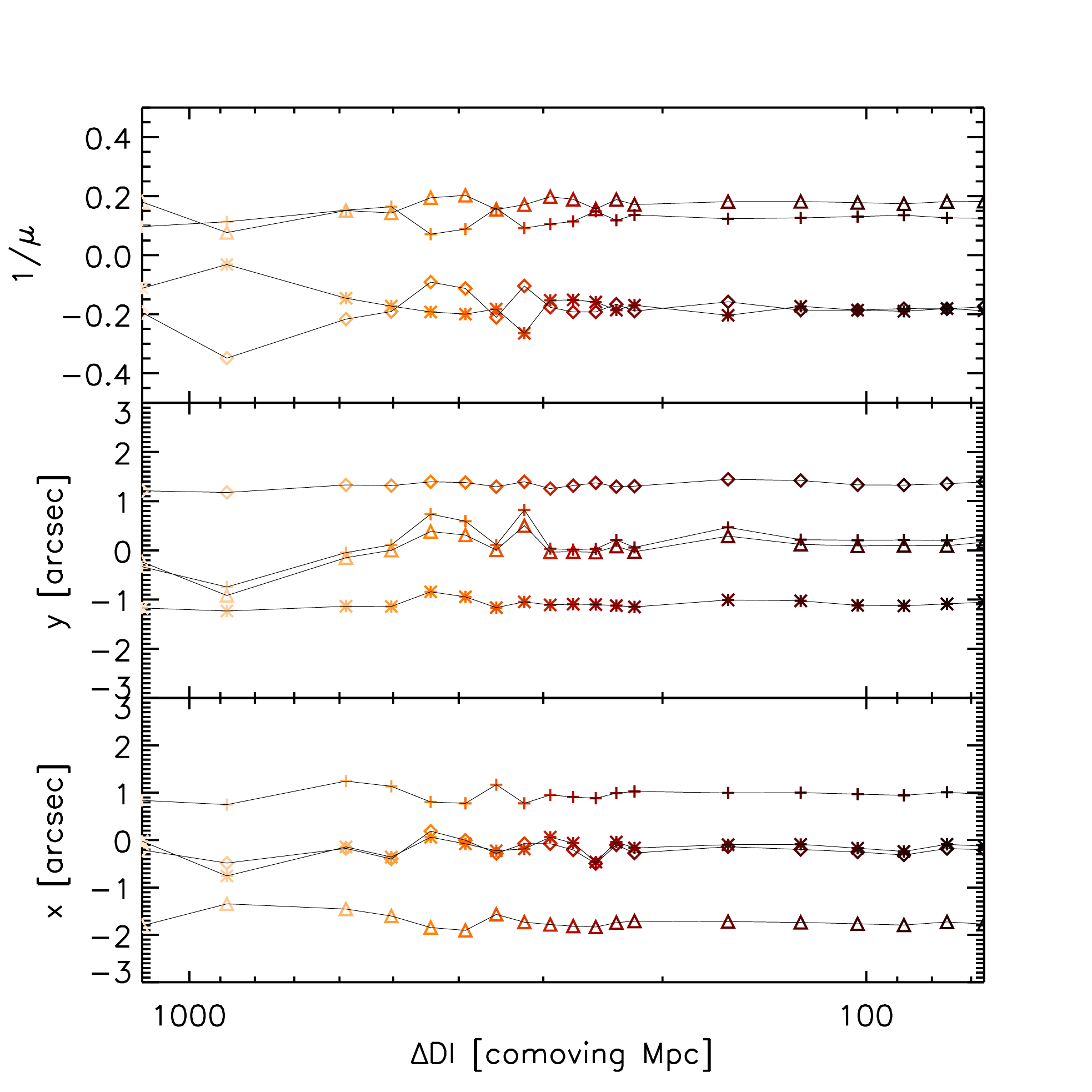}
\caption[]{Image  positions and  magnification  as a  function of  the
  comoving distance  between the  lensing planes (different  number of
  planes) for the lensing simulation  of a quasar at redshift $z=3.62$
  by a  galaxy at redshift  $z=0.42$. Separation of  approximately 300
  \cMpc   \,    corresponds   to    22   lensing   planes    in   this
  case. \label{fig:all}} \end{figure}

It  is important  to note  here that  since we  calculate the  angular
diameter  distances  between   planes  assuming  the  Robertson-Walker
metric, the mass added to those lenses planes will cause our light-cone
to  effectively   contain  more  mass   than  the average  for  the
Friedman-Lemaitre universe used  to calculate the distances  if it is
not compensated  for --  the average angular  size through  the planes
will  the  smaller on  average.   This  is not  a  small  effect if  a
significant fraction of the mass in the universe is represented on the
planes.  When  needed we correct  for this  by add a  uniform negative
mass density to each of the  planes that compensates for the clustered
mass  so that  the  average  surface density  is  zero  on each  plane
\citep{1988ApJ...327..526S}\footnote{This could be  avoided by using a
  Dryer-Roeder  angular diameter  distances.}.  This  will be  further
discussed in section~\ref{sec:light_cones}.

The deflection angles at each plane add to give the angle with respect
to    the    radial   direction    at    the    $i$th   plane.    From
equation~(\ref{eqn:basicM}) it  follows then that the  position on the
$(i+1)$th plane is
\begin{equation}
{\bf x}^{i+1} = {\bf x}^i - D_{i+1,i} \left\{  {\pmb \theta} +
\sum_{j=1}^{i-1} {\pmb \alpha}^j({\bf x}^j) \right\}.
\end{equation}
Applying this to  the $i$th plane in terms of  the $(i-1)$th plane and
subtracting the equations gives a recursion relation for the positions
on the planes
\begin{equation} \label{eq:recursion_x}
{\bf x}^{i+1} = \left( \frac{D_{i+1,i} }{D_{i,i-1}}+ 1 \right){\bf x}^i -
\frac{D_{i+1,i}}{D_{i,i-1}} {\bf x}^{i-1} - D_{i+1,i} {\pmb \alpha}^i({\bf x}^i)
\end{equation}
with the initial conditions
\begin{equation}
{\bf x}^0 = 0~~~~,~~~~{\bf x}^1 = {\pmb \theta} D_1.
\end{equation}
The first  of these is the  requirement that all the  rays converge at
the observer.

The magnification matrix  defined in \eqref{eq:A} can  be expressed on
each plane as
\begin{equation}
{\bf A}^i \equiv \frac{1}{D_i} \frac{\partial {\bf x}^i}{\partial
  {\pmb \theta} }.
\end{equation}
This is what the normal magnification matrix would be if the $(i+1)$th
plane      where      the     source      plane.       Differentiating
equation~\eqref{eq:recursion_x} yields
\begin{align}
{\bf A}^{i+1} = 
\frac{D_{i+1,i-1}}{D_{i,i-1}}  \frac{D_{i}}{D_{i+1}} ~{\bf A}^i
- \frac{D_{i+1,i}}{D_{i,i-1}}  \frac{D_{i-1}}{D_{i+1}} ~{\bf A}^{i-1}  \\
- \frac{D_{i+1,i} D_{i} }{D_{i+1}} ~ {\bf G}^i {\bf A}^i \label{eq:1}
\end{align}
 with the initial conditions
\begin{equation}
{\bf A}^0 = 0 ~~~~~~ {\bf A}^1 = {\bf I}.
\end{equation} 
The forcing term is
\begin{equation}
{\bf G}^i \equiv \frac{\partial {\pmb \alpha}^i}{\partial
  {\bf x}^i } = g_0^i {\bf I} + g_1^i {\pmb \sigma}_1 + g_2^i {\pmb \sigma}_2 .
\end{equation}
where the notation of Appendix~\ref{sec:ellipt-matr-repr} is used.
The coefficients are related to the Newtonian surface potential on
the planes by
\begin{align}\label{eq:shears}
g_0 = \frac{1}{2} \nabla^2\phi ~~~~~~
g_1 = \frac{1}{2} \left(\phi,_{11}
  - \phi,_{22} \right) ~~~~ g_2 = \phi,_{12} 
\end{align}
The term \eqref{eq:1} can be written out as
\begin{align} \label{eq:GA}
{\bf G A} = \left[ g_0 (1-\kappa) - g_1\gamma_1 - g_2\gamma_2 \right]
{\bf I} \nonumber \\
+ \left[ -g_0 \gamma_1+ g_1(1-\kappa) + g_2\gamma_3 \right]
{\pmb \sigma}_1 \nonumber \\
+ \left[ -g_0 \gamma_2+ g_2(1-\kappa) - g_1\gamma_3 \right]
{\pmb \sigma}_2 \nonumber \\
+ \left[ -g_0 \gamma_3- g_1\gamma_2 + g_2\gamma_1 \right]
{\pmb \sigma}_3
\end{align}

It  can be  seen  in  equation~\eqref{eq:GA} that  there  is a  ${\pmb
  \sigma}_3$  component to  the  magnification matrix  which does  not
exist      in      the      case       of      one      lens.       In
appendix~\ref{sec:ellipt-matr-repr}, it  is shown that this  implies a
rotation in  the image  \citep{2006MNRAS.367.1543P}.  The case  of two
lens  planes is  worked  out explicitly  for  pedagogical purposes  in
appendix~\ref{sec:two-plane-lens}.

\section{code implementation}\label{sec:implementation}

Multi-plane lensing is implemented as  an extension to the preexisting
{\small GLAMER} code.  Other  aspects of the {\small GLAMER} code are  described in more
detail  in \cite{M&P2013}  (MP).  The  code is  written in  C++ in  an
object oriented manor so that characteristics of the lenses and sources
can be  chosen by  the user in  a very flexible  fashion.  There  are a
number of options allowed which will be described briefly here.

Rays  are   shot  from  the   observer  to  the  source   plane  using
equations~\eqref{eq:recursion_x} and \eqref{eq:GA}.  In each plane the
deflection,  convergence and  shear are  calculated using  the methods
described  in  MP.   The  mass  distribution  on  each  plane  can  be
represented in several  different ways.  A surface density  map can be
input in  FITS format.  This  option is useful for  representing the
output   of  N-body   simulations   and   semi-analytic  methods   for
constructing   galaxies    and   galaxy   clusters   such    as   {\small MOKA}
\citep{2012MNRAS.421.3343G}.   The   lens  planes  can   also  contain
analytic ``halos''.   The halos can  have a variety of  different mass
profiles -- Navarro-Frenk-White  (NFW) \citep{navarro97}, non-singular
isothermal sphere  (NSIE), power-law,  \cite{hernquist90}, point-mass,
etc.   These  ``halos''  can   represent  the  baryonic  component  of
galaxies, the dark  matter halos or even stars.   The contributions of
the halos  to the  lensing quantities are  calculated using  a modified
tree algorithm when  there are enough of them to  warrant it (see MP).
Finally,  the  mass  on  a  plane can  be  represented  by  simulation
particles  with  an  adaptive  smoothing in  which  case  the  lensing
quantities are calculated by tree  algorithm.  The code allows for any
combination of  these representations.   For example, the  dark matter
halos could be represented by NFW  profiles, the baryonic galaxy by an
NSIE, the mass outside  of halos by a mass map and  the stars by point
masses all at the same time.

Once the  lens has been constructed  rays are shot back  to the source
plane in a uniform grid on the image plane.  This initial gridding can
be used to make shear or  magnification map with uniform resolution if
that is desired.  As described in MP there are a number of options for
refining the  grid of  rays.  The  code can be  made to  find critical
curves and caustics and increase the resolution of them to the desired
level.  It can also be made to refine the grid to resolve a particular
source.  The  ray shooting is  parallelized with POSIX  threads, which
increases the performance significantly  especially when the number of
planes is large.

There are also a variety of options for representing the sources.
The simplest source is circular with uniform surface brightness.  The
source can also have an analytic surface brightness distribution or a
pixelized surface brightness distribution that can be input in FITS
format.  There can be any number of sources and they can be of mixed
type.  Each source can also have a different redshift.  Because of the
adaptive gridding of the rays each source can be resolved by shooting a
relatively small number of rays to the particular redshift of the
source. 


\begin{figure}
\centering
\includegraphics[width=0.53\textwidth]{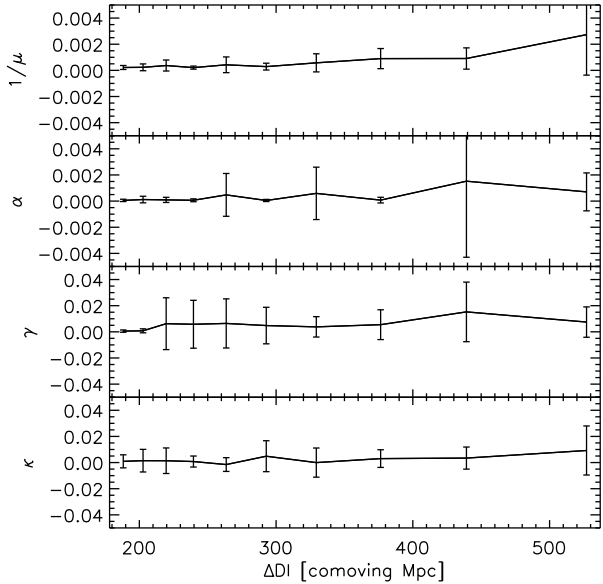}
\caption[]{Relative  error~\eqref{eq:error} of the  lensing properties
  -- inverse  of the magnification (top), magnitude  of the deflection
  angle,  magnitude  of the  shear  and  convergence  (bottom) --  for
  simulations with the same  light-cone, but differing comoving distance between
  the  lensing planes  (different  number of  planes).  Separation  of
  approximately  300 \cMpc \,  corresponds to  18 lensing  planes. The
  field of view  is $10^4 \, \arcsect^2$, the  mass function goes down
  to $10^9\msun$ halos  and up to redshift $z=2.0$.  The true solution
  in this  case is assumed  to be the  one with one plane for each halo 
  at precisely its redshift. \label{fig:mean}} \end{figure}

\begin{figure}
\centering
\includegraphics[width=0.6\textwidth]{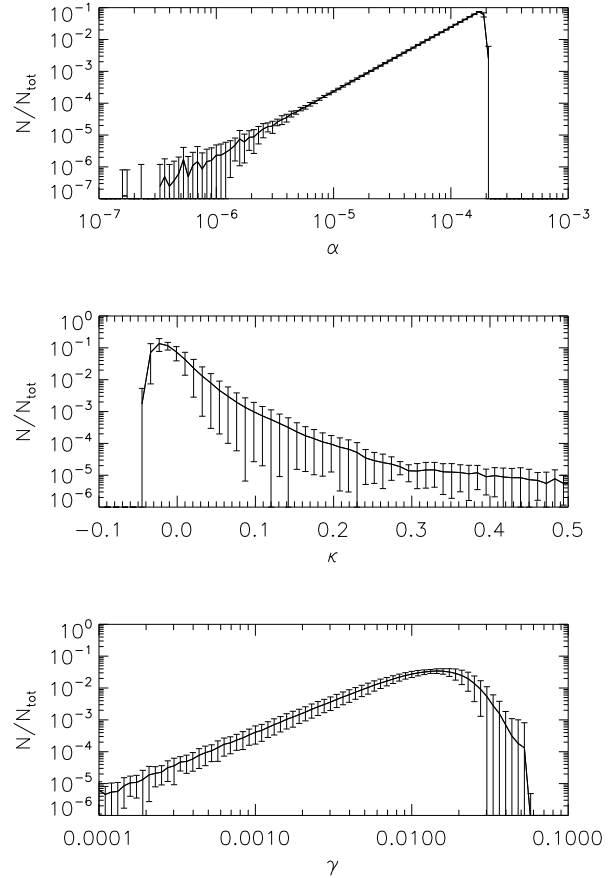}
\caption[]{Distribution  of  the  deflection angle,  convergence,  and
  shear of  a set of  32 random and  uncorrelated fields of  view. The
  area is  $10^4 \, \arcsect^2$ and the  NFW halos are from  a ST mass
  function down  to $10^9\msun$, distributed on 20  planes. The source
  is at redshift $z=2.0$. The results agree very well with simulations
  by \citet{Pace2011} \label{fig:pdf}}
\end{figure}

\section{Generating Light Cones}\label{sec:light_cones}
In this  section, we describe the  different types of light-cones that are
implemented within  or can be  input into  the {\small GLAMER} \,  code.  The
mass within  the light-cone can  be generated in four  different ways:
($i$) particles  can be taken directly  from a N-body simulation  of a
light-cone and  sorted onto planes, ($ii$) halos can  be identified in
the  N-body simulation  and inputted  into {\small GLAMER}  as a  halo catalog,
($iii$) random  halos can be  generated within {\small GLAMER} from  a redshift
dependent  mass  function and  other  analytic  descriptions of  their
internal parameters,  ($iv$) two dimensional maps  of surface density,
which  are  generated  from   a  N-body  simulation  or  semi-analytic
simulation, can be  input into {\small GLAMER}.  A light-cone  can have lensing
planes of any  mixture of these types -- for  example a galaxy cluster
might be represented  by a mass map and the  foreground and background
objects generated randomly from a mass function.  We will discuss each
of these cases below.

\subsection{Light-Cones from Simulation Particles}\label{sec:particle_lc}
{\small GLAMER}  allows the  user to  input a  list of  particle positions  and
smoothing  scales  for a  light -cone.   The  smoothing is  best  done
adaptively according to particle density  in three dimensions -- as in
Smooth Particle  Hydrodynamics (SPH).   The particles are  sorted onto
the  desired number  of  lens  planes and  then  a  tree structure  is
constructed  on  each plane  to  calculate  the deflection  and  other
lensing  quantities  (see  MP   for  details).   For  state-of-the-art
numerical simulations the number of  particles within a complete
light-cone 
can be  very large which makes memory  management difficult.  For
this reason, we favor using methods  ($ii$) or ($iv$) or a combination
of  them to  represent  the output  of a  N-body  simulation, but  the
ability to use the particles  directly is maintained.  A simulation of
a single object with relatively  few particles can be implanted within
a light-cone that is represented in a different way.

\subsection{Light-Cones from Halo Catalog}\label{sec:mill_lc}
{\small GLAMER} \,  is able to read in a catalog of halos and their
properties.  These halos can be taken from an N-body simulation or any
other source.   The halos are projected onto the lensing planes
keeping their angular positions fixed and a tree structure is
constructed on each plane.  The force and deflections are calculated
by tree algorithm (see MP).

To represent  an N-body halo  as an  analytic halo, parameters  of the
halo model must be matched  to quantities calculated from the particle
distribution  --  virial  mass,  half mass  radius,  maximum  circular
velocity,  moments  of  inertia,  etc.   This  procedure  is  somewhat
dependent on the  halo catalog used and what  information it provides.
One can also attempt to represent the baryons that might not be in the
simulation with  an additional ``halo''  within the dark  matter halo.
This requires further  modeling of the relation between  halo mass and
stellar mass  and the  distribution of baryonic  mass within  a N-body
halo of a  given type.  These are interesting subjects  that are being
actively investigated.  Lensing has  great potential for investigating
these relationships.   We will not  discuss them in detail  here since
our purpose is  to introduce a tool for investigating  them and not to
advocate any particular model.

We have implemented one full pipeline  for N-body to sky images using
the Millennium Simulation  catalog of halos and with  sources from the
Millennium Run Observatory  \citep{2013MNRAS.428..778O}.  This project
will be discussed in a separate paper.

\subsection{Light-Cones from Generated Halos}\label{sec:massfunc_lc}

The halos catalog can also be created internally and then sorted onto
lens planes as in the case a catalog from a N-body simulation.   No
attempt has been made yet to cluster these halos or to create subhalos
within halos for the full light-cone.  The halos masses are drawn
randomly from a mass function.   The mass
functions implemented in our code so far are:
\begin{itemize}

\item \citet{press74} (PS);

\item \citet{sheth99b} (ST);

\item Power-Law (PL), given by the following equation:
\begin{eqnarray}
\nonumber \dfrac{\mathrm{d}N}{\mathrm{d}\ln M} &=& \dfrac{A}{M^2} \sqrt{\dfrac{1}{2 \pi}} 
 \left[ \dfrac{M}{M^*(z)} \right]^{\alpha}  \\ 
&\times& \exp \left\{ -\dfrac{1}{2} \left[\dfrac{M}{M^*(z)} (1+z)^{3/2} \right]^{\zeta \alpha}\right\}\,;
\end{eqnarray}
where  $\alpha$  represents  the  slope  of  the  power-law,  $A=0.2$,
$\zeta=1.3$ and  $M^*(z)$ the  non-linear mass  at redshift  $z$ given
such  that  $\delta_c(z)=S(M)$,  with  $\delta_c(z)$  the  overdensity
required for spherical collapse at $z$,  and $S(M)$ is the variance in
the linear fluctuation field when smoothed with a top-hat filter of scale
$R = (3M/4  \pi \bar{\rho})^{1/3}$ where $\bar{\rho}$  is the comoving
density  of the   background.   The \citet{sheth99b} mass function in
$\mathrm{\Lambda}$CD on cluster mass scales is well approximated by
$\alpha=1/6$.
\end{itemize}

The three different mass  function are plotted in Fig.~\ref{fig:massF}
.  For a test, we have  populated light-cones of $10^6 \arcsect^2$ and
measured the mass function in  different redshift bins. The histograms
of  the  halo  masses,  for  which  we  put  together  $64$  different
realizations    of   the    light-cone,    are    also   plotted    in
Fig.~\ref{fig:massF}.  The agreement is very good.

In order  to model the  halo lensing properties  we need to  make some
assumption about the dark matter distribution inside the virial radius
of each system - the virial  radius defines the scale inside which the
overdensity  exceeds   the  critical  value  predicted   for  collapse
\citep{gunn77,bond91,lacey93}.  Numerical simulations predict that the
dark  matter follows  an  NFW  profile \citep{navarro97,navarro04}  or
something very  close to it.   Using this model  we must also  fix the
concentration (the ratio between the scale and the virial radius: $c =
r_s/R$ ) for  each halo.  In {\small GLAMER} \, we  have implemented four model
for the mass-concentration relation (we  refer the reader to the cited
papers  for  more  details):  \citet{zhao09},  \citet{munozcuartas11},
\citet{giocoli12b} and a power-law relation:
\begin{equation}
c =  10 \, \left( \dfrac{M}{10^{12}} \right)^{\alpha} \left[ \dfrac{H_0}{H(z)} \right]^{2/3}
\end{equation}
where the redshift evolution factor is  motivated by the change of the
normalization as seen in cosmological numerical simulations. 

This  method of  populating  the  light-cone  with  halos has  several
limitations  with respect  to using  a halos  catalog derived  from an
simulation.  Since the positioning of  the halos is random, we discard
any clustering or particular environmental effects.  Halos do not have
subhalos within  them.  The advantage  is that there is  no resolution
limit and no limit to how many light-cones can be created.

As  in the  halos  from N-body  case, the  baryonic  component can  be
modeled by adding  an additional halo to the center  of each NFW halo.
So far we  have implemented one simple way of  doing this.  The galaxy
is modeled as a NSIE with a  random orientation.  The mass of the NSIE
is   calculated  using   the  stellar   mass/halo  mass   relation  of
\cite{2010ApJ...710..903M}:
\begin{equation}
M_{\rm star} = m_o \frac{ (M_{\rm halo}/M_1)^{\gamma_1}}{\left[ 1 +
    (M_{\rm halo}/M_1)^\beta\right]^{(\gamma_1-\gamma_2)/\beta}}
\end{equation}
with                                 $m_o=7.3113\times10^{10}\,\msun$,
$M_1=2.8575\times10^{10}\,\msun$,  $\gamma_1=7.17$ ,  $\gamma_2=0.201$
and $\beta=0.557$.   This is  derived from  matching the  abundance of
observed galaxies to the number of  predicted halos in mass bins.  The
NFW halos mass  is reduced accordingly.  Much  more sophisticated ways
of doing this are possible and will be implemented in the future.

The centers  of some halos will  lie outside the light-cone, but they
will  intersect the  cone's  boundary.  This  can  result in  boundary
effects  because  the  density  out  the  boundary  of  the  field  is
artificially low.   A buffering option  is implemented to  avoid this.
All the halos are created in the region outside of the cone but within
a fixed proper distance ($\sim 1 \Mpc$) from the boundary.  The result
is a funnel shape instead of a cone.

\subsection{Mass Maps}\label{sec:mass_map}
Maps of  surface density in  FITS format can  be read into  {\small GLAMER} and
placed on  a lens plane at  the desired redshift.   The deflection and
shear are  found by Fourier transforming the  surface density, solving
equation    \eqref{eq:poissons}   for   the    potential,   evaluating
equations~\eqref{eq:shears} and then Fourier transforming back to real
space.  Maps of  shear and deflection are stored so  that this need be
done  only  once.   The  lensing  quantities  are  then  evaluated  by
interpolating from these maps during the ray shooting procedure.

\begin{figure*}
\centering
\includegraphics[width=1\textwidth]{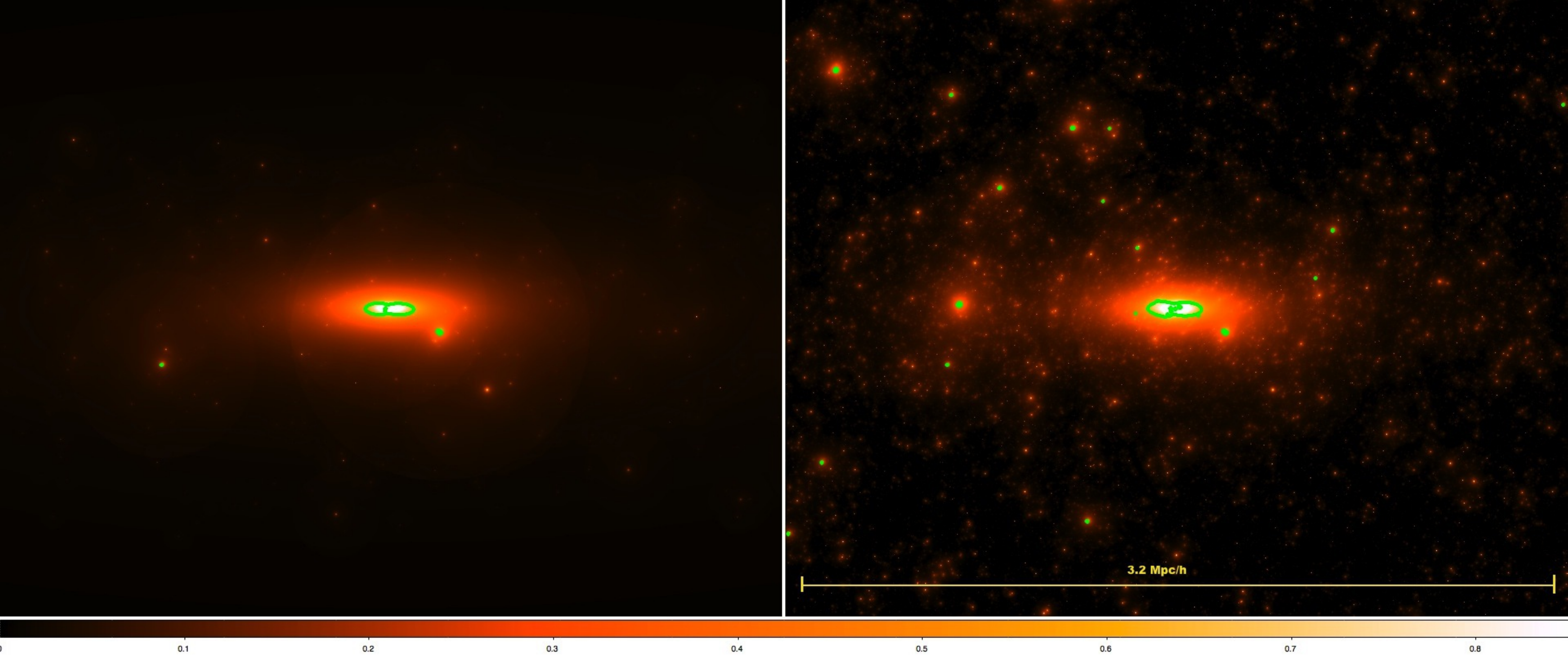}
\caption[]{Convergence  map  of  a  cluster without  (left)  and  with
  (right)  structure along  the line  of sight.   The cluster  lens is
  located at redshift $z=0.3$, it posses a mass of $9.2 \times 10^{14}
  \msun$ and  a concentration of  $7$. The  green contours in  the two
  figures  show  the  location   of  radial  and  tangential  critical
  lines. The cluster on the left  possess an Einstein radius of $18.9$
  arcsec while in the right panel,  where the structure along the line
  of   sight   are   included,   the   Einstein   radius   is   $21.3$
  arcsec. 
\label{fig:cluster_map}}
 \end{figure*}

When using a light-cone taken from a large-scale simulation it is very
advantageous to project the particles in shells around the observer onto
planes and then construct two  dimensional density maps by CIC (clouds
in cell) or some other method.  Because the shells can be quite thick,
as we will  see in section~\ref{sec:tests}, this can  save an enormous
amount  of memory  and  computational time  over  using the  particles
directly without significant lose in accuracy.  We have also used this
capability to implant simulated galaxy  clusters into light-cones that
are otherwise populated  with halos.  Fig.~\ref{fig:cluster_map} shows
and example of this.

\subsection{Sources}
Sources can  be distributed  throughout the light-cone.  They  can be
represented by a number of analytic forms including a circular uniform
surface brightness  profile, a \cite{1963BAAA....6...41S}  profile and
an inclined exponential disk profile.  A pixelized image of the source
can  also be  used.  A  catalog  of sources  can  be read  in from  an
external source or individual sources can be created randomly.  {\small GLAMER}
can adaptively  find and refine  around the  image even when  they are
highly distorted  and there  are multiple images  of the  source.  The
rays are shot only back to the redshift of this source.

A numerical effect can occur if  the source is placed in the center of
a halo and  then the halo is projected to the  nearest lens plane that
is at  lower redshift than  the source.  This will  produce alignments
with significant differences in  radial distance and spurious lensing.
The distance between planes should be small enough to make the lensing
by  the  nearest plane  small,  but  the  perfect alignment  can  have
unwanted effects.  To avoid this, the deflection is turned off for the
lens plane that contains the  source's halo. 

\begin{figure}
\centering
\includegraphics[width=0.54\textwidth]{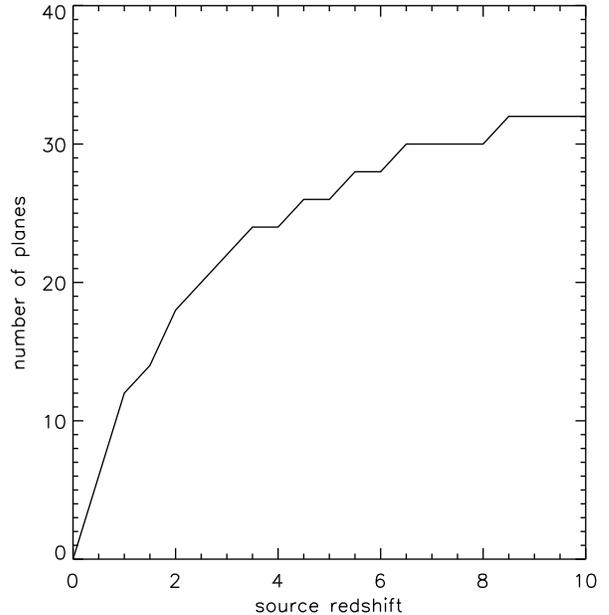}
\caption[]{The number of planes needed in a simulation versus source redshift. The values are chosen  for separation  distance between  lensing planes  of 300
  \cMpc.
\label{fig:np_z}}
\end{figure}

\section{Tests}\label{sec:tests}
We perform several tests to check the accuracy and convergence of our
multi-plane gravitational lensing implementation in the code {\small GLAMER}.
In our first test we determine the
convergence of our code by comparing the  image
positions and magnification of a
four-imaged quasar, as a function of the number of lensing planes.
In Fig.~\ref{fig:pos}  we plot  the position  of the  four images  of the
quasar  at  redshift   $z=3.62$,  lensed  by  a   galaxy  at  redshift
$z=0.42$. The  field of  view of  $10^4 \arcsect^2$  with a  buffer of
1~\Mpc,  filled  with  NFW  halos  from a  ST  mass  function  down  to
$10^9\msun$. We compute the lensing properties and change the number
of planes in two plane interval as the main galaxy lens stays fixed on a
plane at redshift $z=0.42$. 

The  changes in the  image positions are no  greater than
0.4  \arcsect  \, in  one  x-direction  and  1.2  \arcsect \,  in  the
y-direction. This  can also  be observed in  Fig.~\ref{fig:all}, where
plot the image  positions and inverse magnifications as  a function of
the distance between  the planes.  The positions show  very small ($<$
0.2 \arcsect)  scatter for  300  \cMpc  \, or  less,  equivalent to  22
lensing planes. The same holds for the inverse magnification.

Our next test aims to examine how other lensing properties, such  as
shear  and  convergence, converge
with increasing the number of lensing planes. For the study  we
generate a light-cone with a field of view of
$10^4 \, \arcsect^2$ that extends  to redshift $z=2.0$. We populate it
with NFW  halos from  a ST mass function down  to $10^9\msun$.  As a
``true''  result we  consider the  case  where an  infinite number  of
planes  is used,  i.e. each  lensing  halo occupies  a single  lensing
plane. We will denote this results as the ``true'' solution.

We compare the relative error of the lensing properties of simulations
with different  number of planes to  the true solution. To do this we
compare the image maps pixel by pixel and we formulate the 
the relative  error  $\epsilon$ in the following way: 
\begin{equation}  \label{eq:error}\epsilon =  \frac{1}{N_{\rm
    pixels}}  \sum_{i=0}^{N_{\rm pixels}-1}\frac{x_i  - x_i^{\rm true}}{x^{\rm
    true}_i},  \ee  where  \be  N_{\rm  pixels}  =  \sum_{i=0}^{N_{\rm
    pixels}-1} x_i , \end{equation} 
and $x_i$ is  the value of the quantity at the
respective pixel $i$.

In Fig.~\ref{fig:mean} we plot the relative error for the convergence,
shear, deflection  angle, and inverse  magnification as a  function of
the comoving separation between the lensing planes. The relative error
for the deflection  angle and the inverse magnification  is very small
--  for  less than  300  \cMpc  \, between  the  planes  it is  around
0.1\%. The  error for  the convergence and  shear is ten  times larger
 and drops  down to less  than 5\%
for  less than  300 \cMpc,  which  supports the  conclusions from  our
previous test. This separation corresponds  to 18 lensing planes for a
source at redshift $z=2$.

In the next test, we  investigate the agreement of our distributions of
the  lensing quantities  with already  published results.   We compute
histograms of the lensing properties of  a field of view with the same
parameters as the  one from the previous test (field  of view of $10^4
\,  \arcsect^2$, extending  to  redshift $z=2.0$,  populated with  NFW
halos from  a ST mass function  down to $10^9\msun$),  but keeping the
number of planes  fixed to 20. We change  randomly the distribution of
the NFW  halos in the light-cone and  obtain a set of  32 different and
uncorrelated fields of view. In Fig.~\ref{fig:pdf} we show the average
of  the   convergence,  shear,  and   deflection  angle  for   the  32
realizations  of  the  image   plane  and  we  overplot  the  standard
deviation.   The  results  agree   very  well   with  the   ones  from
\citet{Pace2011}, despite  the much  smaller field of  view.

Finally,  we show  how structure  along  the line  of sight  may change  the
Einstein radius of  a galaxy cluster. In our simulation  we define the
Einstein  radius as  the  median distance  of  the tangential  central
critical points from the cluster center \citet{meneghetti10a}.  In the
left panel of Fig.~\ref{fig:cluster_map} we  show convergence map of a
halo  created   with  {\small  MOKA}  \citep{giocoli12}   at  redshift
$z=0.3$.  The cluster  possess a  virial mass  of $9.2  \times 10^{14}
\msun$ and a  concentration of $7$.  In the right  panel we include in
the map the effect of uncorrelated structures along the line of sight,
assuming an  empty light-cone. The  green lines in the  maps represent
the radial and tangential critical  curves, where the magnification of
the background  sources, located at  $z_z=2$, is infinite.   The green
connected  points in  the center  of  the maps  indicate the  Einstein
radius  of  the  cluster,  measured  as the  median  distance  of  the
tangential  central  critical points  to  the  halo center.  From  the
analysis we notice  that while the Einstein radius in  the left figure
is $18.8$ arcsec,  in the right increases to $21.3$  arcsec, where LSS
are  included.  We  will  discuss motivate  and  better quantify  this
effect in a future paper Giocoli et al. in preparation.

We have  also performed further small  test to ensure  the accuracy of
our multi-plane lensing  scheme.  For example, we have  tested that the
average  lensing mass  on each  plane is  zero and that  an empty
light-cone with an arbitrary  number of lensing planes always produces
zero deflection.  We also checked that an  analytic lens placed  in an empty  light-cone
with an arbitrary  number of planes agrees fully  with the single lens
plane calculation.

From our tests we  conclude that a separation of 300  \cMpc \, or less
is sufficient  for accurate  representation of  the lensing  field. In
Fig.~\ref{fig:np_z}  we  show  the correspondence  between  number  of
planes and  source redshift,  that gives  approximately 300  \cMpc \,
plane separation.   For redshift $z=2$,  18 planes are  sufficient. For
much higher redshifts, $z=10$, the number increases to 32.

\section{Some applications}
\label{sec:some-applications}

In this section,  we present some example  applications to demonstrate
{\small GLAMER}'s applicability  to interface with  other lensing codes  and to
produce simulated  observations.  We  do this without  fully describing
the parts  of the  simulations that  are not  directly related  to the
functionality of {\small GLAMER}.

As   previously   discussed,   Fig.~\ref{fig:cluster_map}   shows   the
convergence  map for  a simulated  galaxy cluster.   The cluster  lens
located    at    redshift    $z=0.3$   was    created    using    {\small MOKA}
\citep{2012MNRAS.421.3343G},  and has  as input  parameters a  mass of
$9.2\times 10^{14}M_{\odot}/h$ and a concentration of $7$, and sources
are  located at  redshift $z_s=2$.   The axial  ratios of  the cluster
ellipsoid are  randomly drown  from the  \citet{jing02} distributions,
they are $a/b=0.62$ and $a/c=0.3$ where $a$, $b$ and $c$ represent the
smallest,  intermediate  and  largest  axis  of  the  halo  ellipsoid,
respectively; the random  orientation of the systems,  with respect to
the line of sight, gives in the plane of the sky an axial ratio of the
ellipse that is  $a'/b'=0.32$. As a result the cluster  on the left is
characterized by  an Einstein radius $\theta_E=18.8$  arcsec. In {\small GLAMER}
we  have  created  an  interface   that  allows  {\small MOKA}  outputs  to  be
reprocessed   including   background   and   foreground   objects   by
constructing    a    random    light-cone    as    described    in
section~\ref{sec:massfunc_lc}.  The {\small MOKA} map  was in this way inserted
into the  light-cone.  As in  all these examples, the  ray tracing was
done without  the Born  approximation.  Similar map  of the  shear and
deflection are created.  The critical lines and caustics can be mapped
and the images of lensed background  objects can be found. The created
light-cone realization  modify the  size  of the  Einstein radius  to
$\theta_E=21.3$ arcsec.

\begin{figure*}
\centering
\includegraphics[width=0.8\textwidth]{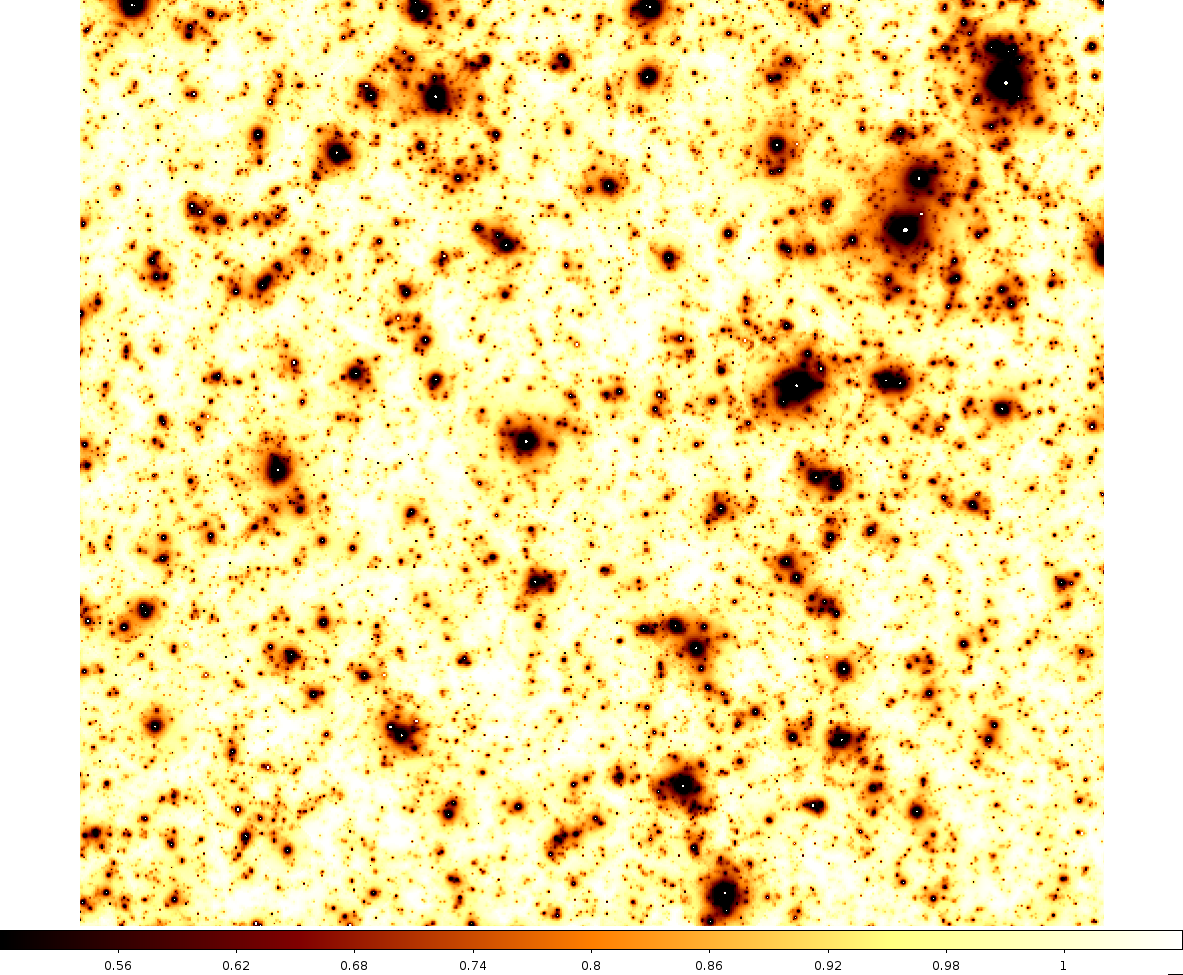}
\caption[]{Map of the  inverse of the magnification,  $1/|\mu|$, for a
  8.4 by 8.4 arcmin random field with the sources plane at z=2.5.  The
  image  is on  a regular  grid,  but the  rays have  been refined  to
  resolve the critical curves so the high magnification ( and the $\mu
  <  0$)  regions   had  higher  resolution  before   this  image  was
  interpolated to a  regular grid.  In some of the  regions where $\mu
  \simeq 1$  the resolution  of the  adaptive grid  is lower  then the
  final image resolution. \label{fig:invmag_map}}\end{figure*}

Fig.~\ref{fig:invmag_map}  shows   a  map   of  the  inverse   of  the
magnification for a random 8.4 by 8.4 arcminute field.  The light-cone
was  constructed from  a halo  catalog extracted  from the  Millennium
simulation.  Each  halo is represented by  a NFW profile for  the dark
matter and  a NSIE profile for  the inner regions.  The  grid in which
the rays have been shot was  refined to give higher angular resolution
where the magnification is high.

\begin{figure*}
\centering
\includegraphics[width=0.8\textwidth]{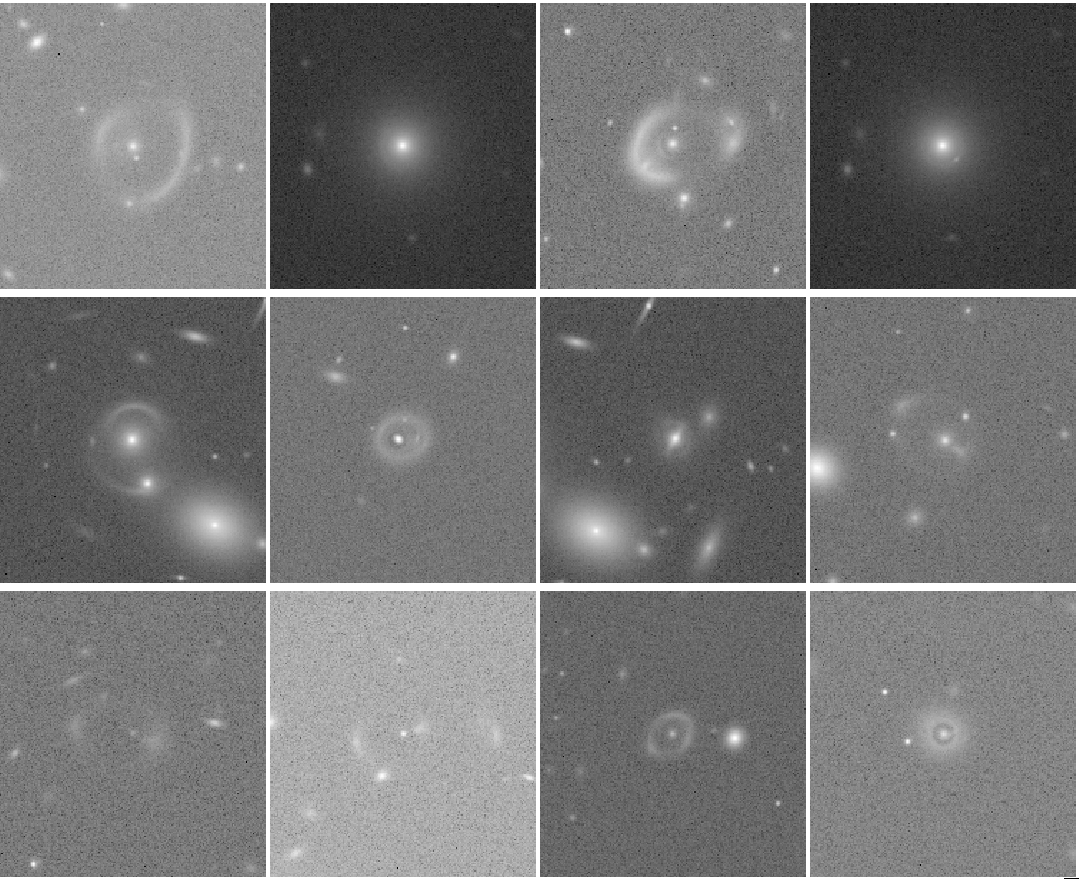}
\caption[]{Twelve simulated galaxy-galaxy lenses. The mass
  distributions of the galaxies and their dark matter halos are taken
  to be NSIE within NFW profiles.  The visible galaxies are modeled
  as \cite{1963BAAA....6...41S} profiles at the centers of the
  halos.  Noise has been added and a Gaussian point spread function
  (psf) applied.  Two of these are the same lens with different
  sources behind it.  The output images are in standard fits files. 
\label{fig:12lens_images}}\end{figure*}

Fig.~\ref{fig:12lens_images}  shows   twelve  examples   of  simulated
galaxy-galaxy  strong lenses.   The ray  shooting grid  is refined  to
adapt to  the surface brightness  of the  images.  Each source  is ray
traced back  to its own  redshift and has  a NFW+NSIE halo.   Here the
caustics in a random light-cone were found and an extra source was put
close to  the caustic.  Many thousands  of such lenses can  be created
easily.  Such  images are being  used to  test lens finding  robots, lens
modeling algorithms and predict the statistics of detectable lenses.

\section{discussion and conclusions}\label{sec:conclusions}
We  have  presented an  extension  of  the ray-shooting  lensing  code
{\small GLAMER} to multiple planes  and its ability to perform lensing
simulations in  a light-cone,  rather than  using a single  lens. This
allows  us  to perform  more  accurate  statistical strong  and  weak
lensing simulations and study properties of the strong lensing galaxies
and galaxy clusters, as well as the structure along the line of sight
outside of the primary lens.

We   have  presented   in   detail  the   methodology  and   numerical
implementation that we have use. The mass in the universe, in the form
of halos,  is projected onto a  discrete number of lensing  planes and
the  lensing  properties,  such  as deflection  and  convergence,  are
computed on each  of these planes and summed up  according to the lens
equations. The summation of the  contributions from the different halos
is performed via a spacial tree algorithm in  order to improve
accuracy and speed.  Details of  the performance and limitations of the
tree can be found in MP.

The halos,  populating a light-cone,  can be either taken  from a dark
matter  simulation  or randomly  generated  from  a mass  function  at
different  redshifts.  For the
second  one  we  have  adopted   three  different  mass  functions  --
Press-Schechter, Sheth- Tormen and a simple power law. In this context
more general  and useful relations  can be implemented and  inputed as
desired.

We have presented various tests and  applications of our code. We have
shown that the optimal number of planes per light-cone is achieved for
a separation  of circa 300  \cMpc.  We  have presented that  the image
position of  a lensed  quasar by  a galaxy  converges for  this value.
Moreover, the relative uncertainty between the lensing properties in a
simulation with that  given plane separation and  an infinitesimal one
is  less  than   3\%.   We  have  also  shown   that  the  statistical
distribution for the lensing properties  in our simulations agree very
well with previously published results.
Finally, we have demonstrated some of the applications that the code is
already being applied to.  

\section*{Acknowledgments}
This research is part of the project GLENCO, funded
under the Seventh Framework Programme, Ideas, Grant
Agreement n. 259349.

\bibliographystyle{mn2e}


\appendix
\section{ellipticity in tensor representation} 
\label{sec:ellipt-matr-repr}

We have found the following formalism  useful in dealing with propagating magnification matrices and dealing with galaxy ellipticities. 

The magnification matrix, ellipticity or any 2-by-2 matrix can be represented as
\begin{equation}
{\bf E} =  \rho {\bf I} + \epsilon_1 {\pmb \sigma_1} + \epsilon_2 {\pmb \sigma_2} + \epsilon_3 {\pmb \sigma_3}.
\end{equation}
where $\epsilon_1$, $\epsilon_2$ and $\epsilon_3$ are scalar
coefficients which may depend on position on the sky.  The matrices are
\begin{align}
{\bf I}  = 
\left( 
\begin{array}{rr}
1 & 0 \\
0 & 1
\end{array}
\right)~~~~~~~~~~~~~~~~ \\
~~~~~~~{\pmb \sigma_1}  = 
\left( 
\begin{array}{rr}
-1 & 0 \\
0 & 1
\end{array}
\right) ~~~~~
{\pmb \sigma_2}  = 
\left( 
\begin{array}{rr}
0 & -1 \\
-1 & 0
\end{array}
\right) \\
{\pmb \sigma_3}  = 
\left( 
\begin{array}{rr}
0 & 1 \\
-1 & 0
\end{array}
\right)~~~~~~~~~~~~~~
\end{align}
These matrices obey the following algebraic relations 
\begin{align}
\begin{array}{lll}
{\pmb \sigma_1}{\pmb \sigma_2} = {\pmb \sigma_3} 
&{\pmb \sigma_1}{\pmb \sigma_3} = {\pmb \sigma_2} 
&{\pmb \sigma_3}{\pmb \sigma_2} = {\pmb \sigma_1} \\
{\pmb \sigma_1}{\pmb \sigma_1} = {\bf I} 
&{\pmb \sigma_2}{\pmb \sigma_2} = {\bf I} 
&{\pmb \sigma_3}{\pmb \sigma_3} = - {\bf I} 
\end{array} \label{eq:algebra}
\\
{\pmb \sigma_i}{\bf I} = {\bf I}{\pmb \sigma_i} = {\pmb \sigma_i} 
~~,~~{\pmb \sigma_i}{\pmb \sigma_j} = - {\pmb \sigma_j}{\pmb \sigma_i}  
\mbox{  ,  for } i\neq j \nonumber
\end{align}
Some of the standard quantities and transformations are easily expressed in this basis:
\begin{align}
& \trace {\bf E} = 2\,\rho 
& \mbox{trace} \label{trace}\\
&\det {\bf E} = | {\bf E} | = \rho^2 - \epsilon_1^2 - \epsilon_2^2 + \epsilon_3^2 & \mbox{determinant} \label{determinant}\\
&{\bf E}^{\rm T} = \rho {\bf I} + \epsilon_1 {\pmb \sigma_1} + \epsilon_2 {\pmb \sigma_2} - \epsilon_3 {\pmb \sigma_3} 
& \mbox{transpose} \label{transpose}\\
&{\bf E}^{-1} = \frac{1}{| {\bf E} |}\left[ \rho {\bf I} - \epsilon_1 {\pmb \sigma_1} - \epsilon_2 {\pmb \sigma_2} - \epsilon_3 {\pmb \sigma_3} \right]
& \mbox{inverse}. \label{inverse}
\end{align}

The rotation operator in this space is
\begin{equation}
{\bf R} = \cos(\phi) {\bf I} - \sin(\phi) {\pmb \sigma_3}
\end{equation}
Using the algebraic relations, the behavior of these matrices under
rotations can be written
\begin{equation} \label{eq:rotsigma1}
{\bf R}^{-1}{\pmb \sigma_1}{\bf R} = \cos(2\phi)  {\pmb \sigma_1} - \sin(2\phi)  {\pmb \sigma_2}
\end{equation}
\begin{equation}
{\bf R}^{-1}{\pmb \sigma_2}{\bf R} = \cos(2\phi)  {\pmb \sigma_2} + \sin(2\phi)  {\pmb \sigma_1}
\end{equation}
\begin{equation}
{\bf R}^{-1}{\pmb \sigma_3}{\bf R} =   {\pmb \sigma_3} 
\end{equation}
Thus ${\bf I}$ and ${\pmb \sigma_3}$ are rotationally invariant.

The ellipticity of a galaxy can be represented by a matrix with only
${\pmb \sigma_1}$ and ${\pmb \sigma_2}$ components
\begin{equation}
{\bf \hat{E}} = \epsilon_1 {\pmb \sigma_1} + \epsilon_2 {\pmb \sigma_2}.
\end{equation}
 Using equation~(\ref{eq:rotsigma1}) it can be seen that an ellipticity
with only an $\epsilon_1$ component will become an ellipticity with a $\epsilon_2$ 
component when rotated.  In this way an angle can be associated with
any ellipticity -- the rotation angle required to rotate the
ellipticity into having only an $\epsilon_1$ component.  This angle is
defined only up to an additive integer multiple of $\pi$.

We can multiply ellipticities together using the algebraic
relations \eqref{eq:algebra} 
\begin{align}
{\bf \hat{E}}^{(1)} {\bf \hat{E}}^{(2)}    =&
 \left( \epsilon_1^{(1)}\epsilon_1^{(2)} 
  +  \epsilon_2^{(1)}\epsilon_2^{(2)} \right) {\bf I}  \nonumber \\
&~~ +  \left( \epsilon_1^{(1)}\epsilon_2^{(2)}
  -  \epsilon_2^{(1)}\epsilon_1^{(2)} \right)  {\pmb \sigma_3} \\
 =&~ |\epsilon^{(1)}| |\epsilon^{(2)}| \left[\cos\left( 2(\phi^{(1)} - \phi^{(2)}) \right)
   {\bf I}  \right.  \nonumber \\
& ~~~~~~~~~~~\left. +   \sin\left( 2(\phi^{(1)} - \phi^{(2)}) \right) {\pmb
     \sigma_3} \right]
\end{align}
where $|\epsilon| = \epsilon_1^2 + \epsilon_2^2$.   This product
is rotationally invariant and describe the relative orientation of the
ellipticities.  Half its trace can be used as a scalar product.


\section{Two plane lens}
\label{sec:two-plane-lens}

To get a better intuitive feel for what is happening we can apply the equations in section~\ref{sec:multip_formalism} to the case
where there are two lens planes.  In this case, the position on the
source plane  ($i = 3$) is
\begin{align}
{\bf x}^3 = & ~ D_3 {\pmb \theta} - 
D_{3,1} {\pmb  \alpha}^1\left[D_1 {\pmb \theta}\right] \nonumber \\
& - D_{3,2} {\pmb \alpha}^2\left[D_2 {\pmb \theta} 
- D_{2,1}  {\pmb  \alpha}^1\left[D_1 {\pmb \theta}\right] \right]
\end{align}
and the magnification is
\begin{align}
{\bf A}^3 = & ~ {\bf I} - \frac{D_{3,2}D_2}{D_3} {\bf G}^2 -
\frac{D_{3,1}D_1}{D_3} {\bf G}^1 \label{eq:lnearterms}\\
& + \frac{D_{3,2} D_{2,1} D_1}{D_3} {\bf G}^2 {\bf G}^1.\label{eq:nonlinearterm}
\end{align}
The nonlinear term written out is
\begin{align}
{\bf G}^2 {\bf G}^1 &= \left( g_o^1g_o^2 +  g_1^1g_1^2 +  g_2^1g_2^2
\right) {\bf I}   \nonumber \\
&+  \left( g_o^1g_1^2 +  g_1^1g_o^2 \right) {\pmb \sigma}_1 
+  \left( g_o^1g_2^2 +  g_2^1g_o^2 \right) {\pmb \sigma}_2   \nonumber
\\
&+  \left( g_1^1g_2^2-g_2^1g_1^2 \right) {\pmb \sigma}_3.
\end{align}
It is seen in these formulae that the magnification matrices of each of the lens planes acting alone add (\ref{eq:lnearterms}), but there is an additional nonlinear contribution (\ref{eq:nonlinearterm}) which gives rise to a rotation.

\section{transformation of the moments}

For completeness we include here the transformation of an image's
moments by the gravitational lensing distortion in the tensor
formalism.  We feel that this is a more intuitive way to express these
transformations than the more common ways in the literature \citep[see][]{1991ApJ...380....1M,KS93,bartelmann01}.

The moments tensor of an image with surface brightness ${\mathcal I}({\pmb \theta})$ can be define as
\begin{align}
{\bf M} =& \int d^2\theta ~ {\mathcal I}({\pmb \theta}) ~ ({\pmb \theta} - {\pmb \theta}_o ) ( {\pmb \theta} - {\pmb \theta}_o)^{\rm T} \\
=& ~\rho {\bf I} + \epsilon_1 {\pmb \sigma}_1 + \epsilon_2 {\pmb \sigma}_2.
\end{align}
where ${\pmb \theta}_o$ is the centroid of the image.
The surface brightness is conserved but the coordinates are transformed
\begin{align}
{\bf M} & = \int d^2\theta ~ {\mathcal I}({\pmb \theta}) ~ (\delta{\pmb \theta} ) ( \delta{\pmb \theta})^{\rm T} \\
& =  \int \frac{d^2\beta}{|{\bf A}|} ~ {\mathcal I}({\pmb \beta}) ~ {\bf A}^{-1}\delta{\pmb \beta}  ( {\bf A}^{-1}\delta{\pmb \beta})^{\rm T} \\
& =  \int \frac{d^2\beta}{|{\bf A}|} ~ {\mathcal I}({\pmb \beta}) ~ {\bf A}^{-1}\delta{\pmb \beta}\delta{\pmb \beta}^{\rm T}  ( {\bf A}^{-1})^{\rm T}\\
& = \frac{1}{|{\bf A}|}  {\bf A}^{-1} {\bf M}'  ( {\bf A}^{-1})^{\rm T}
\end{align}
where it is assumed that the magnification matrix is constant over the source.
Primes will indicate quantities for the original, pre-lensed source.  Written out in our tensor notation this is
\begin{align}
 {\bf M}  = &  \frac{1}{|{\bf A}|} &\left[ \rho'  {\bf A}^{-1} ( {\bf A}^{-1})^{\rm T} + \epsilon'_1~ {\bf A}^{-1} {\pmb \sigma}_1 ( {\bf A}^{-1})^{\rm T} \right. \\
& & \left.+  \epsilon'_2 ~ {\bf A}^{-1} {\pmb \sigma}_2 ( {\bf A}^{-1})^{\rm T} \right].
\end{align}
This can be readily worked out in terms of convergence and shear using
equations~(\ref{determinant}), (\ref{transpose}) and (\ref{inverse})
along with the decomposition of ${\bf A}$~\eqref{eq:Adecomposition} (note the negative signs),
\begin{align} \label{eq:moment_transform}
 {\bf M}  = & \frac{(1-\kappa)^2}{|{\bf A}|^3}  
& \left\{   \rho' \left[ \right. \right.&
\left( 1 + \hat{\gamma}_1^2 + \hat{\gamma}_2^2 + \hat{\gamma}_3^2 \right){\bf I} \\
& & &-2\left( \hat{\gamma}_1 + \hat{\gamma}_2 \hat{\gamma}_3  \right){\pmb \sigma}_1 \nonumber \\
& & &\left. -2\left( \hat{\gamma}_2 - \hat{\gamma}_1\hat{\gamma}_3 \right){\pmb \sigma}_2 \right] \nonumber \\
& &+  \epsilon_1' \left[ \right. &-2 \left( \hat{\gamma}_1 + \hat{\gamma}_2 \hat{\gamma}_3  \right)
{\bf I} \nonumber \\
& & & + \left( 1 + \hat{\gamma}_1^2 - \hat{\gamma}_2^2 - \hat{\gamma}_3^2 \right) {\pmb \sigma}_1 \nonumber \\
& & &\left. + 2 \left( \hat{\gamma}_3 + \hat{\gamma}_1\hat{\gamma}_2 \right){\pmb \sigma}_2 \right] \nonumber \\
& & +  \epsilon_2' \left[ \right. &-2 \left(  \hat{\gamma}_2 - \hat{\gamma}_1 \hat{\gamma}_3  \right)
{\bf I}  \nonumber \\
& & & - 2 \left( \hat{\gamma}_3 - \hat{\gamma}_1\hat{\gamma}_2 \right) {\pmb \sigma}_1 \nonumber \\
& & &\left.\left. + \left( 1 - \hat{\gamma}_1^2 + \hat{\gamma}_2^2 - \hat{\gamma}_3^2 \right) {\pmb \sigma}_2 \right] \nonumber \right\}
\end{align}
where $\hat{\gamma}$ is the reduced shear,
\begin{align}
\hat{\gamma}_i \equiv \frac{\gamma_i}{(1-\kappa)}.
\end{align}

Equation~\eqref{eq:moment_transform} is the full transformation
without averaging over the unknown orientation of the source or making
the weak lensing approximation (while it is assumed that the shear and convergence are constant over the whole image).  Averaging over the orientation of the source gives
 $\langle \epsilon'_1 \rangle = |\epsilon'| \langle \cos(2\phi) \rangle = 0$
and, by the same argument, $\langle \epsilon'_2 \rangle=\langle
\epsilon'_1 \epsilon'_2 \rangle = 0$ and $\langle \rho' \epsilon'_1
\rangle = \langle\rho' \epsilon'_2 \rangle = 0$.  By expanding to
first order in shear and convergence (the weak lensing approximation)
and then averaging over the source's orientation one obtains,
\begin{align}
\left\langle \frac{\bf M}{{\rm tr} {\bf M}} \right\rangle \simeq &   \frac{1}{2} {\bf I} -\hat{\gamma}_1 {\pmb\sigma}_1 - \hat{\gamma}_2 {\pmb\sigma}_2 \\
& +  \left\langle {\epsilon'_1}^2  \right\rangle \hat{\gamma}_1    {\pmb \sigma}_1 
 +   \left\langle {\epsilon'_2}^2  \right\rangle \hat{\gamma}_2   {\pmb \sigma}_2 \\
= & \frac{1}{2} {\bf I} - \left( 1-\frac{|\epsilon'|^2}{2\rho'^2} \right)\left( \hat{\gamma}_1 {\pmb\sigma}_1 + \hat{\gamma}_2 {\pmb\sigma}_2 \right)
\end{align}
where $\langle {\epsilon'_1}^2\rangle = \langle {\epsilon'_2}^2\rangle = |
\epsilon' |/2$ has been used.  For convenience and to comply with the
common definition of ellipticity, we define $\hat{\epsilon}_1 =
\epsilon_1/\rho$ and $\hat{\epsilon}_2 =
\epsilon_2/\rho$. \footnote{It is
  common to define the components of ellipticity as $e_1 =
  ({\bf M}_{11} -{\bf  M}_{22})/{\rm tr}{\bf M}$ and $e_2 = 2
  {\bf M}_{12}/{\rm tr}{\bf M}$.}

It is convenient to renormalize moments matrix and subtract the trace
\begin{align}
 {\bf Q}  \equiv \frac{1}{1-\langle|\hat{\epsilon}'|^2\rangle/2} \left( \frac{1}{2} {\bf I} -   \frac{\bf M}{\trace {\bf M}} \right)
\end{align}
where the average in the denominator is now over both the magnitude
and the orientation on the source ellipticity.  When ${\bf Q}$ is
average in this way we get
\begin{align}
\left\langle {\bf Q} \right\rangle  \simeq 
   \gamma_1  {\pmb \sigma}_1 + \gamma_2 {\pmb \sigma}_2.
\end{align}
Thus ${\bf Q}$ is an estimator of the weak shear.  We find this
expression simpler and more intuitive than the more common
transformations using ``polars'' \citep{KS93} or complex ellipticities \citep{1997A&A...318..687S}.
 
\bsp

\end{document}